\documentclass[usenatbib,useAMS]{mn2e}
\usepackage{graphicx}
\usepackage{bm,amssymb,amsmath}
\usepackage{times}
\usepackage{natbib}
\usepackage{float}
\usepackage{ulem}
\usepackage{color}
\usepackage{fixltx2e} 
\usepackage{ulem} 
\graphicspath{{./ }{./png/}{./ps/}}

\newcommand{\vect}[1]{{{\mbox{\boldmath $#1$}}}}		

\newcommand\average[1]{{\langle #1 \rangle}} 		
\newcommand\averaged[1]{{\langle #1 \rangle}} 		
  \newcommand{\phm}{{\phantom{0}}} 	
    \newcommand{\turb}{_{\rm t}}     			
  \newcommand{\dd}{{\rm d}}       				

\newcommand{\corr}{{L}}

  \newcommand{\cm}{\,{\rm cm}}

  \newcommand{\erg}{\,{\rm erg}}

  \newcommand{\kms}{\,{\rm km\,s^{-1}}}

  \newcommand{\kpc}{\,{\rm kpc}}
  \newcommand{\pc}{\,{\rm pc}}
  
  \newcommand{\Myr}{\,{\rm Myr}}
  \newcommand{\Gyr}{\,{\rm Gyr}}
  
  \newcommand{\mkG}{\,\mu{\rm G}}
  \newcommand{\nG}{\,{\rm nG}}

  \newcommand{\p}{\,{\rm pc}}

  \newcommand{\HDI}{{Paper~I}} 				

  \newenvironment{frcorr}
    {\color{black}}{}

\pagerange{\pageref{firstpage}--\pageref{lastpage}}
\label{firstpage}
\pubyear{2012}

\begin{document}

\title[SN-regulated ISM. II]{The supernova-regulated ISM. II. 
The mean magnetic field}
  \author[Gent et al.]{
    F.~A.~Gent,$^1$\thanks{
      E-mails: F.A.Gent@ncl.ac.uk,
      Anvar.Shukurov@ncl.ac.uk,
      G.R.Sarson@ncl.ac.uk,
      Andrew.Flet\-cher@ncl.ac.uk and
      Maarit.Mantere@helsinki.fi
      }
    A.~Shukurov$^1$,
    G.~R.~Sarson$^1$,
    A.~Fletcher$^1$,
    M.~J.~Mantere$^2$\\
    $^1$School of Mathematics and Statistics,
      Newcastle University, Newcastle upon Tyne NE1~7RU, UK\\
    $^2$Physics Department, 
    University of Helsinki, PO BOX 64, Helsinki, FI-00014, Finland
    }

  
    \date{Received  2012 November 27; in original form 2012 June 28. Accepted
    2012 November 28. }

  \maketitle

  \begin{abstract}
  The origin and structure of the magnetic fields in the interstellar medium of 
  spiral galaxies is investigated with 3D, non-ideal,
  compressible MHD simulations, including stratification in the galactic
  gravity field, differential rotation and radiative cooling.
  A rectangular domain, 
$1\times1\times2\kpc^3$ in size,  
 spans both sides of the
  galactic mid-plane.
  Supernova explosions drive transonic turbulence.
  A seed magnetic field grows exponentially to reach a statistically  
  steady state within 1.6\,Gyr.
  Following \citet{G92} 
we use volume averaging
with a Gaussian kernel to 
  separate magnetic field into a mean field and fluctuations.
  Such 
averaging does not satisfy all Reynolds rules,
  yet allows a formulation of mean-field theory. The mean field
  thus obtained varies in both space and time. 
  Growth rates differ for the
  mean-field and fluctuating field and there is clear scale
  separation
  between the two elements,
  whose integral scales are about
  $0.7\kpc$ and $0.3\kpc$, respectively.
  \end{abstract}


  \section{Introduction}\label{Intro}
  The interstellar medium (ISM) of spiral galaxies is strongly affected by  
  energy injection from supernovae (SNe),
  which drive highly compressible, transonic turbulent motions.
  This makes it extremely inhomogeneous,
  yet it supports magnetic fields at a global scale of a few kiloparsecs.
  Mean-field dynamo models have proved successful in modelling galactic 
  magnetic fields and offer a useful framework to study them and to interpret
  their observations \citep[e.g.,][]{BBMSS96,S07}. 
  Turbulent dynamo action involves two distinct mechanisms.   
  The \textit{fluctuation dynamo\/} relies solely on the random nature of the
  fluid flow to produce \textit{random\/} magnetic fields at scales smaller
  than the integral scale of the random motions. 
  The \textit{mean-field\/} dynamo produces magnetic field at a scale 
  significantly larger than the integral scale, and requires rotation and
  stratification to do so.
  For any dynamo mechanism, it is important to distinguish the 
  \textit{kinematic\/} stage when magnetic field grows exponentially as it
  is too weak to affect fluid motions, and the \textit{nonlinear\/} stage when
  the growth is saturated, and the system settles to a statistically steady 
  state.  

  The scale of the mean field produced by the dynamo is controlled by the
  properties of the fluid flow. 
  For example, in the simplest $\alpha^2$-dynamo in a homogeneous, infinite
  domain, the most rapidly growing mode of the mean magnetic field has 
  scale of order $4\pi\eta\turb/\alpha$, where $\alpha$ can be understood as 
  the helical part of the random velocity and $\eta\turb$ is the turbulent
  magnetic diffusivity \citep{SSR83}. 
  For any given $\alpha$ and $\eta\turb$, 
  this scale is finite and the mean field produced is, of course, not uniform. 

  In galaxies, the helical turbulent motions and differential rotation
  drive the so-called $\alpha\omega$-dynamo, where the mean field has a
  radial scale of order $\Delta r\simeq3|{\cal{D}}|^{-1/3}(hR)^{1/2}$ at the
  kinematic stage \citep{SS89}, with ${\cal{D}}$ the dynamo number, 
  $h\simeq0.5\kpc$ the half-thickness of the dynamo-active layer and
  $R\simeq3\kpc$ the scale of the radial variation of the local dynamo 
  number. 
  For $|{\cal{D}}|=20$, this yields $\Delta r\simeq1.3\kpc$.

  These estimates refer to the most rapidly growing mode of the mean magnetic 
  field in the kinematic dynamo; it can be accompanied by higher modes that
  have a more complicated structure.
  Magnetic field in the saturated state can be even more inhomogeneous due to
  the local nature of dynamo saturation.
  The mean magnetic field can have a nontrivial, three-dimensional
  spatial structure, and any analysis of global magnetic structures must 
  start with the separation of the mean and random (fluctuating) parts.
  However, many numerical studies of mean-field dynamos define the
  mean magnetic field as a \textit{uniform\/} field obtained 
by averaging over the whole volume available,
  or in the case of fields showing non-trivial
  variations in a certain direction, as planar averages, e.g., over
  horizontal planes for systems that show $z$-dependent fields
  \citep[e.g.,][]{BS05}.

  The mean and random magnetic fields are assumed to be separated
  by a scale, $\lambda$, of 
  order the integral scale of the random motions, $l_0$;
  $\lambda$ is not necessarily precisely equal to $l_0$, however,
  and must be determined
  separately for specific dynamo systems.  The leading large-scale
  dynamo eigenmodes themselves have extended Fourier spectra, both 
  at the kinematic stage and after distortions by the
  dynamo nonlinearity.  Thus, both the mean and random magnetic fields
  are expected to have a broad range of scales, and
  their spectra can overlap in wavenumber space.  Thus, it is
  important to develop a procedure to isolate a mean magnetic field
  without unphysical constraints on its spectral content.  This
  problem is especially demanding in the multi-phase ISM, where the
  extreme inhomogeneity of the system can complicate the spatial
  structure of the mean magnetic field.

  The definition of the mean field as a horizontal average may be appropriate 
  in simplified numerical models where the vertical component of the mean 
  magnetic field, $\average{B_z}$, vanishes because of periodic boundary
  conditions applied in $x$ and $y$; otherwise,
  $\nabla\cdot\average{\vect{B}}=0$ cannot be ensured.
  An alternative averaging procedure that retains three-dimensional spatial
  structure within the averaged quantities is volume averaging with a 
  kernel $G_\ell(\vect{r}-\vect{r'})$, 
  where $\ell$ is the averaging length: $\average{f}_\ell=\int_V  f(\vect{r}')
  G_\ell(\vect{r}-\vect{r'})\,d^3\vect{r}'$, for a random field $f$. 
  A difficulty with 
volume
averaging, appreciated early in the
  development of turbulence theory, is that it does not obey the Reynolds
  rules of the mean (unless $\ell\to\infty$),
  $\average{\average{f}_\ell g}_\ell\neq\average{f}_\ell\average{g}_\ell$, and 
  $\average{\average{f}_\ell}_\ell\neq\average{f}_\ell$
  \citep[Sect.~3.1 in][]{MY07}.
  Horizontal averaging represents a special case with $\ell\to\infty$ in
  two dimensions, and thus satisfies the Reynolds rules; 
  however, the associated loss of a large part of the spatial structure 
  of the mean field limits its usefulness.
  \citet{G92} suggested a consistent approach to 
volume
averaging which does
  not rely on the Reynolds rules.
  A clear, systematic discussion of these ideas is provided by
  \citet[][Chapter~2]{E12}, and an example of their application can be found in
  \citet{E05}.
  The averaged Navier--Stokes and induction equations remain
  unaltered, independent of the averaging used, if the mean
  Reynolds stresses and the mean electromotive force are defined in 
an appropriate,
  generalized way. The equations for the fluctuations naturally change,
  and care must be taken for their correct formulation.
  An important advantage of averaging with a Gaussian kernel 
  (Gaussian smoothing) is its similarity to astronomical
  observations, where such smoothing arises from the finite width of a
  Gaussian beam, or is applied during data reduction.

  Here we analyze magnetic field $\vect{B}$ produced by the rotational shear
  and random motions in the numerical model of the SN-driven ISM presented by
  \citet[][hereafter, \HDI]{FG12}, using 
Gaussian smoothing.
  We suggest an approach to determine the appropriate length $\ell$, and then
  obtain the mean magnetic field $\vect{B}_\ell$.
  The procedure ensures that $\nabla\cdot\vect{B}_\ell=0$. 
  The random magnetic field $\vect{b}{\frcorr{_\ell}}$ is then obtained as
  $\vect{b}{\frcorr{_\ell}}=\vect{B}-\vect{B}_\ell$.
  The Fourier spectra of the mean and random magnetic fields overlap in 
  wavenumber space, but their maxima are well separated.

  \section{Basic equations}\label{sect:NI}

  We model the ISM, with parameters generally typical of the Solar 
  neighbourhood, using a three-dimensional Cartesian grid in a region 
  $1.024\times1.024\times 2.176\kpc^{3}$ in size, with $1.024\kpc$ in the
  radial and azimuthal directions and $1.088 \kpc$ vertically on either
  side of the galactic plane.
  $D=1.024\kpc$ is the largest effective length scale resolved in all
  directions.
  The injection of thermal and kinetic energy by SNe heats the gas and drives 
  intense random motions, whose correlation length is of order $l_0\simeq100\p$
  within $\pm200\p$ of the galactic plane (\HDI).
  This part of the computational domain encompasses of order 400 turbulent
  cells, so statistical properties of the ISM can be determined reliably.

  Our code (based on the {\sc Pencil Code}, 
  http://code.google. com/p/pencil-code/) and model have been tested to ensure
  that the relevant physical processes, including the expansion of individual
  SN remnants, are modelled reliably (\HDI).
  The minimum numerical grid spacing required for that is $\Delta=4\p$, and we
  use $256\times256\times544$ mesh points (excluding boundary zones), with the
  $x,y,z$ coordinates corresponding to the $r,\phi,z$ coordinates of a
  rotating cylindrical reference frame whose $z$-axis is aligned with the 
  angular velocity of galactic rotation and whose $x$-axis points away from
  the galactic centre.

  \label{subsec:eqs}

  The basic equations include the mass conservation equation, the 
  Navier--Stokes equation, the heat equation --- all presented in \HDI, but
  now including additional terms due to the interaction with the magnetic
  field --- and the induction equation written in terms of the vector 
  potential $\vect{A}$ (in the gauge $\Phi=\eta\nabla\cdot\vect{A}$):
  \begin{align}
    \label{eq:ind}
    {\frcorr{   \frac{\partial \vect{A}}{\partial t}                              }}=&\mbox{}\phm{\frcorr{\vect{u}\times\vect{B}-S A_y\vect{\hat{x}}-S x \frac{\partial \vect{A}}{\partial y} \, + }}\\ \nonumber
     &\mbox{}\phm(\eta+\zeta_\eta)\nabla^2\vect{A}
      +(\nabla\cdot\vect{A})\, ({\frcorr{\nabla\eta+}}\nabla\zeta_\eta),
  \end{align}
  where $\vect{u}$ is the deviation of the gas velocity from the background
  shear profile $\vect{U}=(0,Sx,0)$,
   $\vect{B}$ is the magnetic flux density, and $\eta$ is the
  magnetic diffusivity.
  The angular velocity in the Solar vicinity is $\Omega_0=-S=25\kms\kpc^{-1}$,
  but here we consider a model with $\Omega=2\Omega_0=-S$ to enhance the
  mean-field dynamo action \citep[][]{Gressel08} and thereby save computation
  time.
  The system is driven by localized injection of kinetic and thermal energy
  modelling SN explosions.
  To control widespread strong shocks, we include shock-capturing transport
  coefficients, including $\zeta_\eta$, which differ from zero in converging 
  flows.
  We 
use here
a power-law parameterization of radiative cooling labelled WSW in 
  Table~1 of \HDI. 

  We  apply periodic boundary conditions in the azimuthal ($y$) direction.
  Differential rotation is modelled using the shearing-sheet approximation with
  sliding periodic boundary conditions in $x$.
  Boundary conditions on the top and bottom faces allow gas to escape
  without preventing inward flows, and
  for the magnetic field
we adopt
${\partial A_x}/{\partial z}={\partial A_y}/{\partial z}=A_z=0\Rightarrow B_x=B_y=0$.
Vertical magnetic energy flux
vanishes and 
periodic boundaries restrict the planar averages of $B_z$ to 
be
zero.
The volume or planar averages of $B_x$ and $B_y$ are unrestricted and may 
generate \textit{uniform} 
contributions to
the horizontal magnetic field.

  The initial conditions, described in \HDI, are close to hydrostatic and
  thermal equilibrium.
  We include a weak initial azimuthal magnetic field with 
$B_y=0.03\mkG$
  at the mid-plane, decreasing with $z$ in proportion to gas density $\rho$, so 
  that the initial field averaged over the whole domain is about $0.1\nG$.
  In 200--400\Myr, the hydrodynamic parameters of the system settle into a 
  quasi-stationary turbulent state independent on the initial conditions. 

  \section{The mean magnetic field}\label{sect:dyn}\label{subsec:mb}

  \begin{figure}
  \centering
  \includegraphics[width=0.85\columnwidth]{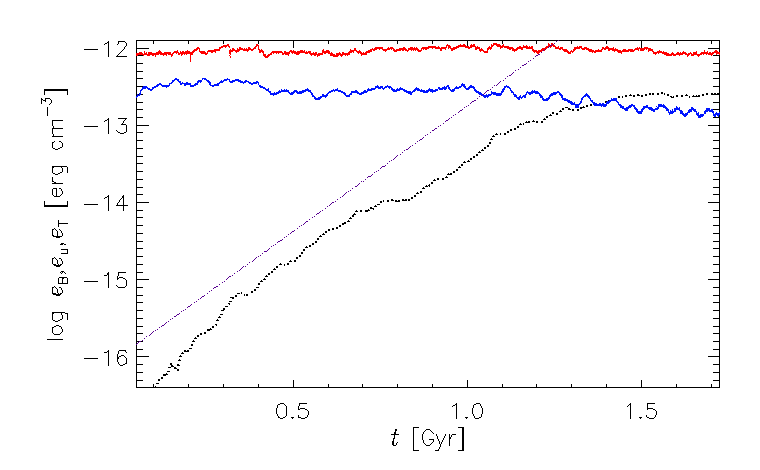}
  \caption{
  Evolution of energy densities 
averaged over the whole
  computational domain; thermal $\averaged{e_{\mathrm{T}}}$ (red), kinetic
  $\averaged{e_{\mathrm{u}}}$ (blue) and magnetic $\averaged{e_{\mathrm{B}}}$
  (black).
  Guide line (purple, dotted) indicates $\exp(7.5t\Gyr^{-1})$.
  \label{fig:bbelog}}
  \end{figure}

  Figure~\ref{fig:bbelog} demonstrates the nearly exponential growth of
  magnetic energy density against the (approximately) stationary background of
  turbulent motions and thermal structure.
  As the average magnetic energy density 
  $\averaged{e_{\mathrm B}}=\averaged{|\vect{B}|^2}/8\pi$ becomes comparable to
  the average kinetic energy density $\averaged{e_{\mathrm{u}}}$ at $t>1\Gyr$, 
  the latter shows a modest reduction, as expected for the conversion of
  the kinetic energy of random motions to magnetic energy.
  For $t>1.4\Gyr$, magnetic field settles to a statistically steady state
  $\averaged{e_{\mathrm B}}\approx2.5\times10^{-13}\erg\cm^{-3}$, somewhat
  larger than $\averaged{e_{\mathrm{u}}}\approx1.6\times10^{-13}\erg\cm^{-3}$,
  while 
the thermal energy density
  $\averaged{e_\mathrm{T}}\approx10^{-12}\erg\cm^{-3}$ 
  {\frcorr{appears only weakly affected. Due to changes to the flow and 
      thermodynamic composition, Ohmic heating is offset by reduced viscous 
      heating or more efficient radiative cooling.}}

  Magnetic field $\vect{B}$ can be decomposed into $\vect{B}_\ell$, the part
  averaged 
over the length scale
$\ell$, 
and the complementary fluctuations $\vect{b}{\frcorr{_\ell}}$,
  \begin{equation} \label{eq:meanB}
    \vect{B}=\vect{B}_\ell+\vect{b}{\frcorr{_\ell}},
    \quad 
    \vect{B}_\ell=\average{\vect{B}}_\ell,
    \quad 
    \vect{b}{\frcorr{_\ell}}=\vect{B}-\vect{B}_\ell,
  \end{equation}
  using 
volume averaging with a Gaussian kernel:
  \begin{align}\label{eq:Bxgauss}
  \average{ \vect{B}}_\ell(\vect{x})
	&=\int_{V}\vect{B}(\vect{x}')G_\ell(\vect{x}-\vect{x}')\,\dd^3\vect{x}',\\
    G_\ell(\vect{x})&=\left(2\pi \ell^2\right)^{-{3}/{2}}
	\exp\left[-{\vect{x}^2}/({2\,\ell^2})\right],\nonumber
  \end{align}
  where $V$ is the volume of the computational domain.
  This operation preserves the solenoidality of both $\vect{B}_\ell$ and 
  $\vect{b}{\frcorr{_\ell}}$ and retains their three-dimensional structure.
  For computational efficiency, the averaging was performed in the Fourier space
  where the convolution of Eq.~\eqref{eq:Bxgauss} reduces to the product
  of Fourier transforms.
  
  \begin{figure}
  \centering\vspace{-0.25cm}
  \includegraphics[width=0.85\linewidth]{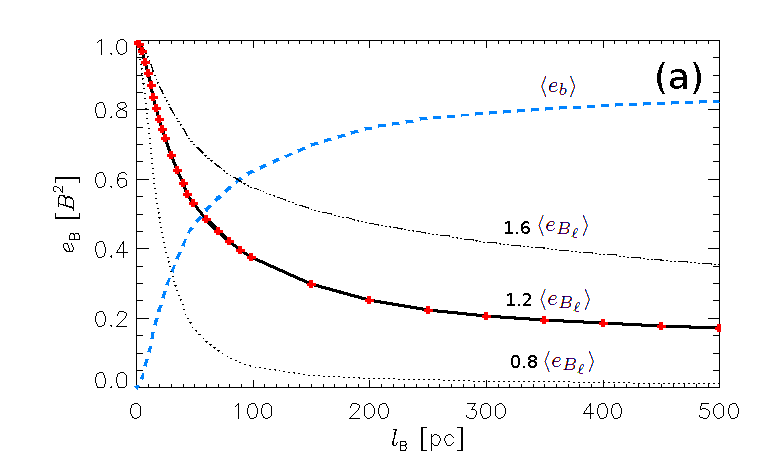}\vspace{-0.2cm}
  \includegraphics[width=0.85\linewidth]{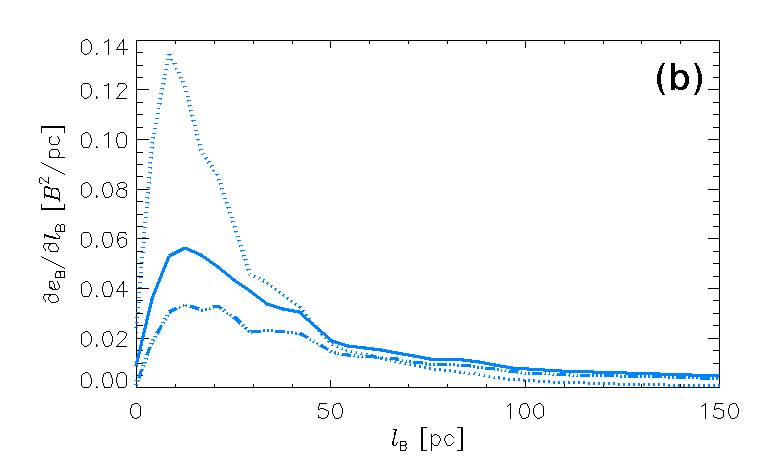}\vspace{-0.2cm}
  \caption{
  \textbf{(a)} Energy densities at $t=1.2\Gyr$ of mean,   
  $\averaged{e_{B_\ell}}$ (black, solid), and fluctuating, 
  $\averaged{e_b}$ (blue, dashed), magnetic fields 
  as functions of the averaging length $\ell$, normalized by 
  $\averaged{\averaged{e_{\mathrm B}}_\ell}$ and averaged over the region $|z|<0.5\kpc$; also
  $\averaged{e_{B_\ell}}$ at $t=0.8\Gyr$ (dotted) and $t=1.6\Gyr$ 
  (dash-{\frcorr{triple}}-dotted). 
  Values of $\ell$ sampled are indicated by red crosses.
  \textbf{(b)}~Derivative of $\averaged{e_b}$ from Panel (a) with respect to
  $\ell$, 
with
the same line types.
  \label{fig:blb}}
  \end{figure}

  Since the averaging \eqref{eq:Bxgauss} does not obey the Reynolds rules for 
  the mean, the definitions of various averaged quantities should be 
  generalized as suggested by \citet{G92}.
  In particular, the local energy density of the fluctuation field is
  given by
  \begin{equation}\label{energyb}
  e_b(\vect{x})=\frac{1}{8\pi}\int_{V}
	|\vect{B}(\vect{x}')-\vect{B}_\ell(\vect{x})|^2
	G_\ell(\vect{x}-\vect{x}')\,\dd^3\vect{x}'.
  \end{equation}
  This ensures that $\average{e_B}_\ell=e_{B_\ell}+e_{b}$,
  where 
  $e_B=|\vect{B}|^2/(8\pi)$ 
and $e_{B_\ell}=|\vect{B}_{\ell}|^2/(8\pi)$.
  Expanding $\vect{B}(\vect{x}')$ in a Taylor series around $\vect{x}$ and 
  using $\int_{V}G_\ell(\vect{x}-\vect{x}')\,\dd^3\vect{x}'=1$ (normalization)
  and $\int_{V} (\vect{x}-\vect{x}')G_\ell(\vect{x}-\vect{x}')\, 
  \dd^3\vect{x}'=0$ (symmetry of the kernel), it can be shown that
  $e_b=|\vect{b}{\frcorr{_\ell}}|^2/(8\pi)+O(\ell/L)^2$, where $L$ is the scale of the averaged
  magnetic field defined as $L^2=|\vect{B}_\ell|/|\nabla^2\vect{B}_\ell|$ in
  terms of the characteristic magnitude of $\vect{B}_\ell$ and its second
  derivatives.
  Thus the difference between the ensemble and volume averages rapidly 
  decreases as the averaging length increases, $\ell/L\to0$, 
  or if the averaging is performed over the whole space, $L\to\infty$;
  this quantifies the deviations from the Reynolds rules
for finite $\ell/L$.

  The appropriate choice of $\ell$ is not obvious. We consider a range
  $0<\ell<500\p$, applying the averaging to 37 snapshots of the magnetic field 
  between $t=0.8$ and 1.7\,Gyr;  the results for $t=0.8$, 1.2 and $1.6\Gyr$ are
  shown in Fig.~\ref{fig:blb}. 
  The smaller is $\ell$, the closer the correspondence between the averaged
  field and the original field (since the average is effectively sampling
  a smaller local volume), and hence the smaller the part of the total field 
  considered as the fluctuation. Hence $\averaged{B_\ell^2}$ is a
  monotonically decreasing function of $\ell$, whereas $\averaged{b{\frcorr{_\ell}}^2}$
  monotonically increases (Fig.~\ref{fig:blb}a). 
  Our aim is to identify that value of $\ell$ where the variation of
  $\vect{B}_\ell$ and $\vect{b}{\frcorr{_\ell}}$ with $\ell$ becomes weak enough.
  To facilitate this, we consider the rate of change of the
  relevant quantities with $\ell$, shown in Fig.~\ref{fig:blb}b (note the
  different scale of the horizontal axis in this panel). 
  The length $\ell\approx50\p$ is clearly distinguished:
  all the curves in Panel~(b) are rather featureless for $\ell>50\p$. 
  The values in Fig.~\ref{fig:blb} have been obtained from the part of the
  domain with $|z|<0.5\kpc$, where most of the gas resides and where dynamo 
  action is expected to be most intense.
  The results, however, remain quite similar if the whole computational
  domain $|z|\leq1\kpc$ is used.
  While this value of $\ell$ has been estimated in a rather heuristic 
  and approximate manner,
  the analysis of the magnetic power spectra in 
  Section~\ref{subsec:scales} confirms
  that $\ell=50\p$ is close to the optimal choice.

  \begin{figure*}
  \centering
  \includegraphics[width=0.3765\linewidth]{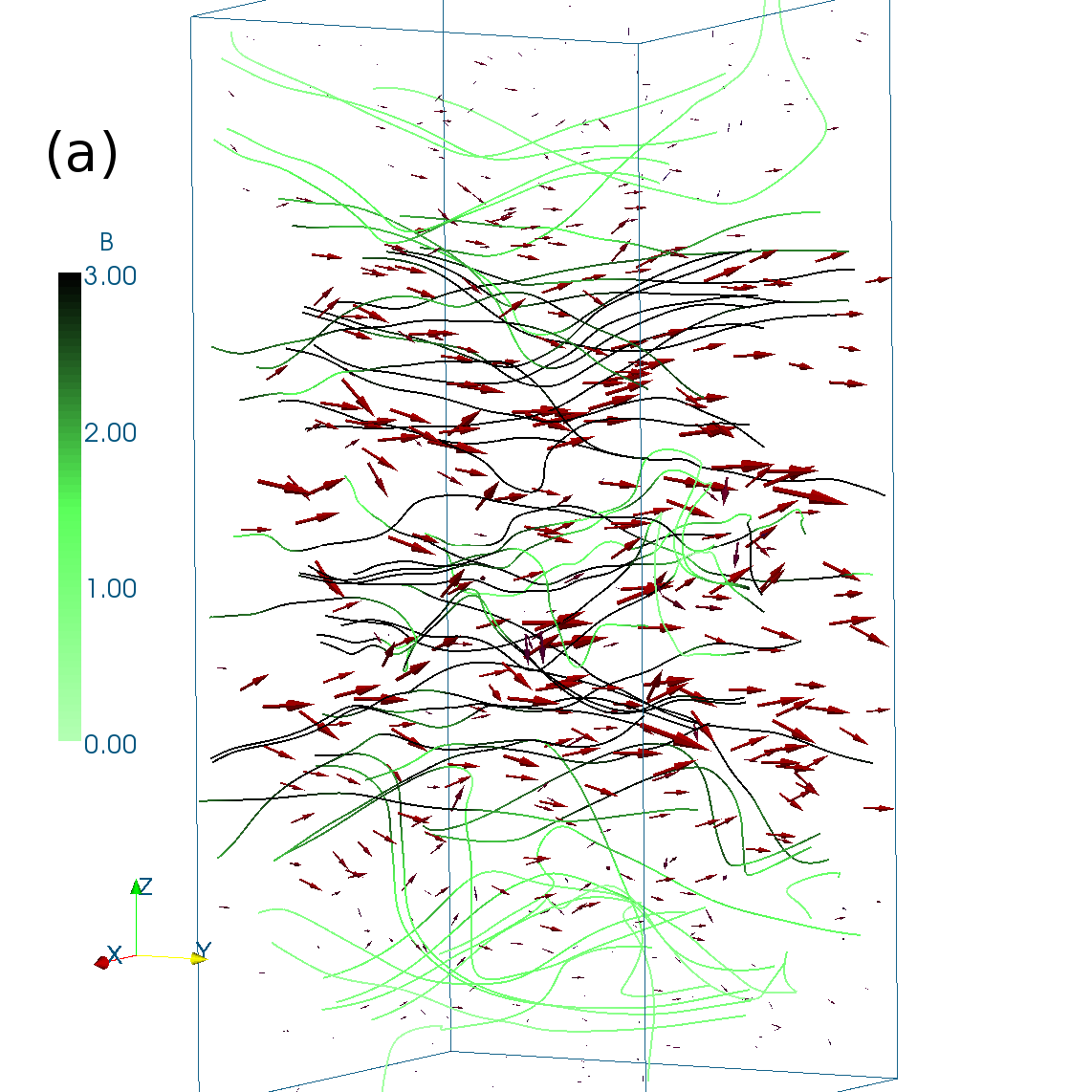}\hspace{-1.25cm}
  \includegraphics[width=0.3765\linewidth]{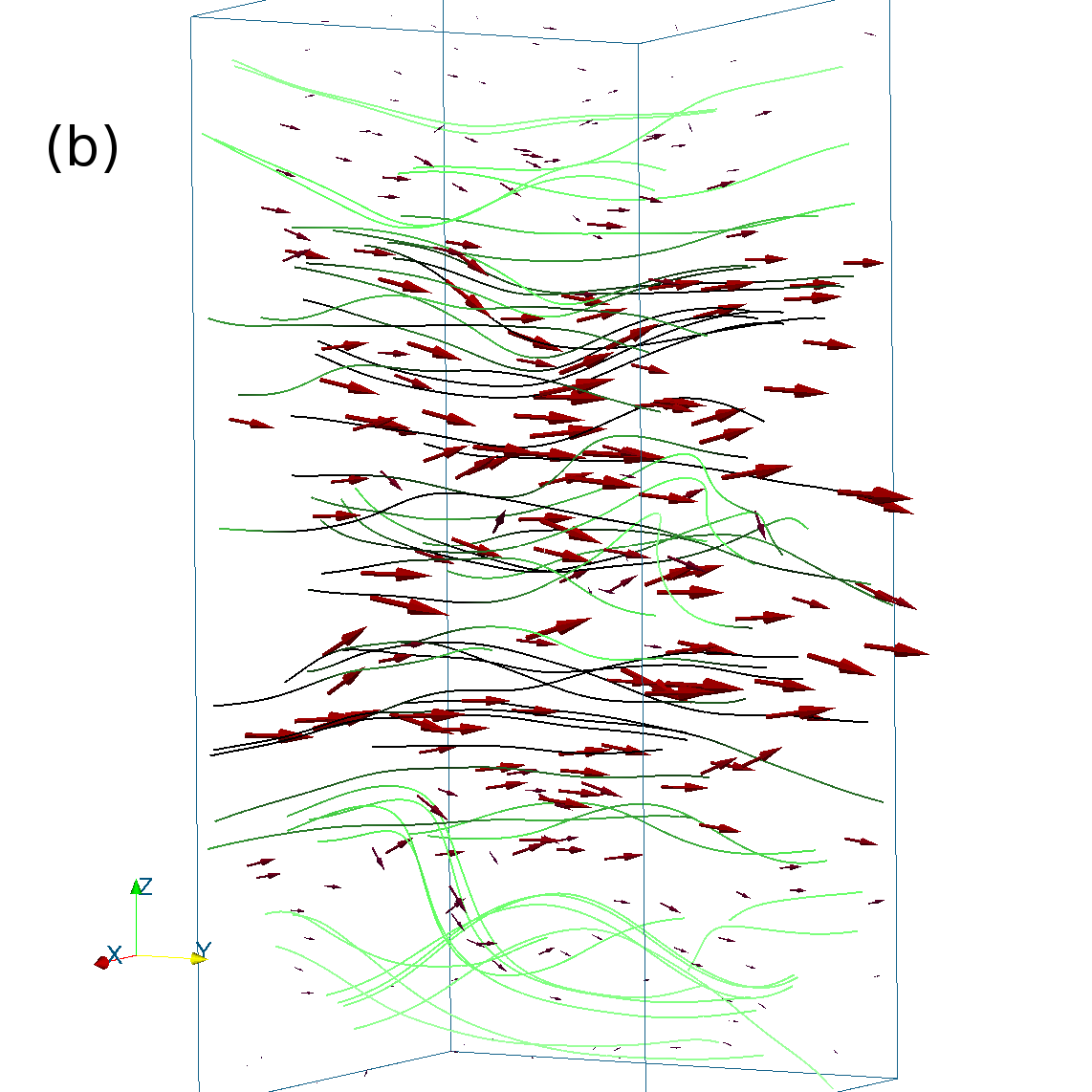}\hspace{-1.25cm}
  \includegraphics[width=0.3765\linewidth]{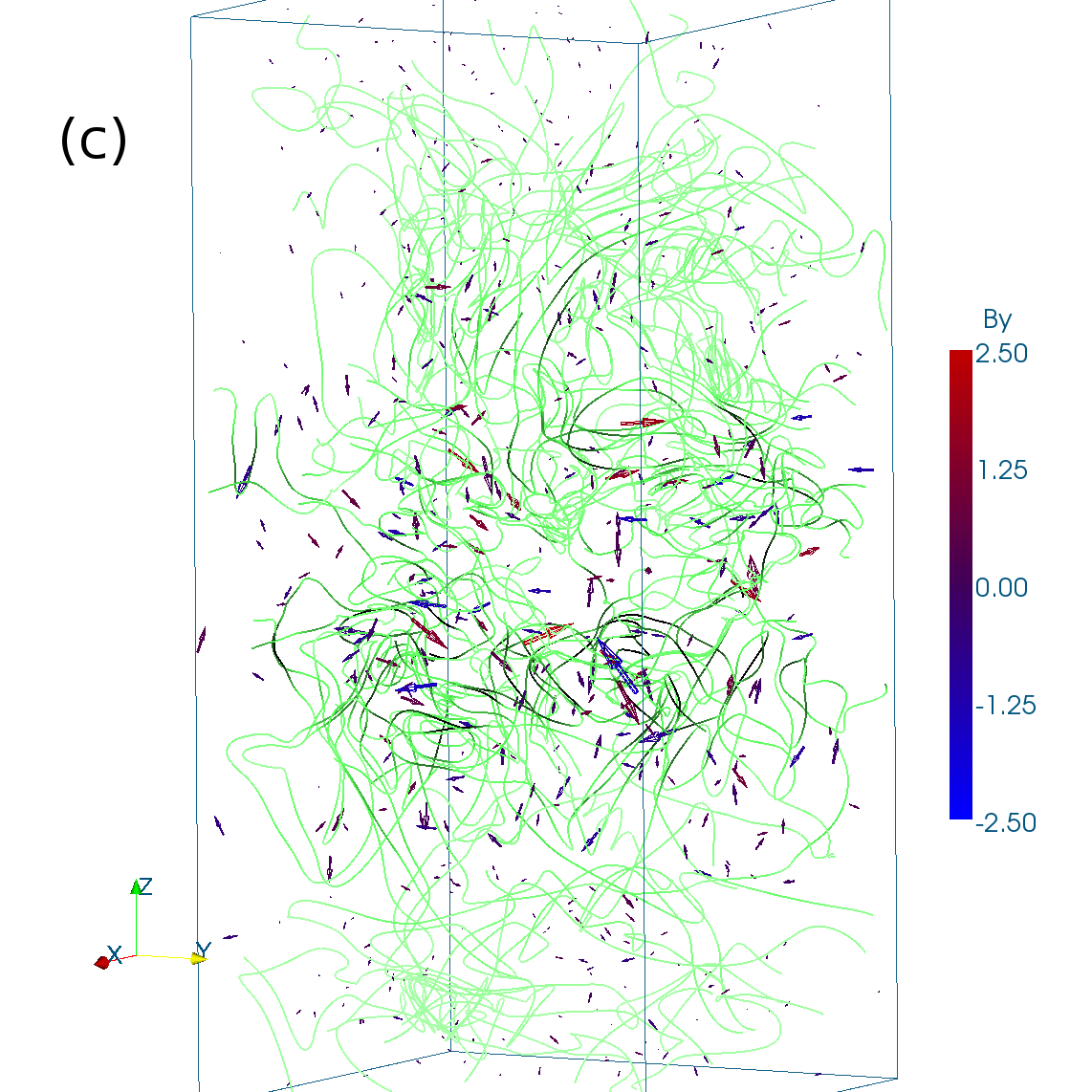}
  \caption{
  Field lines of (a) the total magnetic field $\vect{B}$, 
  (b) its averaged part $\vect{B}_\ell$, (c) the fluctuations $\vect{b}{\frcorr{_\ell}}$,
  obtained by averaging with $\ell=50\p$, for $t=1.625\Gyr$.
  Field directions are indicated by arrows.
  The colour of the field lines indicates the field strength 
  (colour bar on the left), 
  whereas the vectors are coloured according to the strength of the
  azimuthal ($y$) component (colour bar on the right).
  \label{fig:box}}
  \end{figure*}

  To put the estimate $\ell\approx50\p$ into context, we note that it
  is about half the integral scale of the random motions, $l_0$, 
  in a similar model of {\HDI}.
  Simulations of a single SN remnant in an ambient density similar to that at
  $z=0$, reported in \HDI, also show that the expansion speed of the remnant
  reduces to the ambient speed of sound when its radius is 50--70\,pc. 

  Figure~\ref{fig:box} illustrates the total, mean and random magnetic fields thus
  obtained.
  For this saturated state, the field has a very strong uniform
  azimuthal component and a weaker radial component.
The orientation of the field is the same  above and below the mid-plane
($B_y>0$ and $B_x<0$), with maxima located at $|z|\approx0.2\kpc$;
the results will be reported in detail elsewhere.

  \section{Scale separation}{\label{subsec:scales}
  Scale separation between the mean and random magnetic fields in natural 
and simulated turbulent dynamos {\frcorr{can be difficult to identify}}. 
  The signature of scale separation sought for is a pronounced minimum in
  the magnetic power spectrum at an intermediate scale, larger than the 
  energy-range scale of the random flow and smaller than the size of the 
  computational domain.
  {\frcorr{The power spectrum for $\vect{B}$ is}} 
  $ {\frcorr{M(k)=k^{-2}{\average{|\mathcal{F}({k})|}_k},}}$ 
  {\frcorr{for spherical shells of 
thickness $\delta k$ at radius $k=|\vect{k}|$, from}}
      $ {\frcorr{\mathcal{F}(\vect{k})
    ={\int_V \vect{B}(\vect{x})\exp({-2\pi i\vect{k}\cdot\vect{x}})\dd^3\vect{x}}.}}$
  The spectra of the mean and random magnetic fields obtained by 
Gaussian smoothing, 
shown 
in Fig.~\ref{fig:ffta},
have maxima at significantly different wavenumbers.
  We note that the spectrum of the total magnetic field does not have any
  noticeable local minima and, with the standard approach (e.g., based on
  horizontal averages), the system would be considered to lack scale 
  separation. 
     
  \begin{figure}
  \centering\vspace{-0.25cm}
  \includegraphics[width=0.75\linewidth]{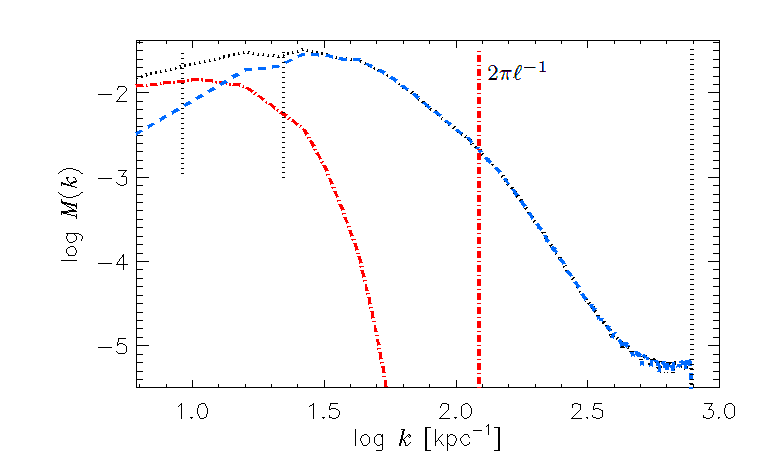}
  \caption{
  Power spectra of total (black, dotted), mean (red, dash-dotted) 
  and fluctuating (blue, dashed) magnetic fields, for 
Gaussian smoothing  
  with $\ell=50\p$, at $t=1.05\Gyr$.  
  Vertical red (dash-dotted) line marks the averaging wavenumber 
  $2\pi\ell^{-1}$ and vertical black (dotted) line indicates the Nyquist wave
  number $\pi\Delta^{-1}$.
  Short vertical segments mark the energy-range scales 
  $2\pi\corr_\ell^{-1}$ and 
  $2\pi\corr_b^{-1}$ of the mean and random magnetic fields, respectively,
obtained from Eq.~\eqref{Bcorr}.
  \label{fig:ffta}}
  \end{figure}

  \begin{figure}
  \centering\vspace{-0.25cm}
  \includegraphics[width=0.75\linewidth]{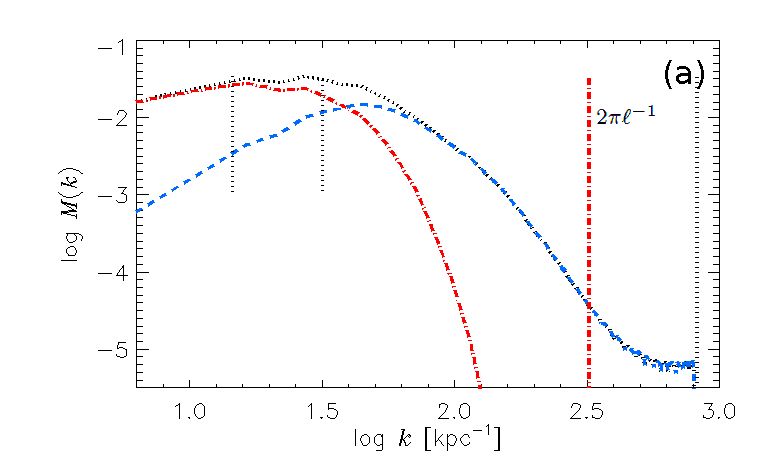}\vspace{-0.2cm}
  \includegraphics[width=0.75\linewidth]{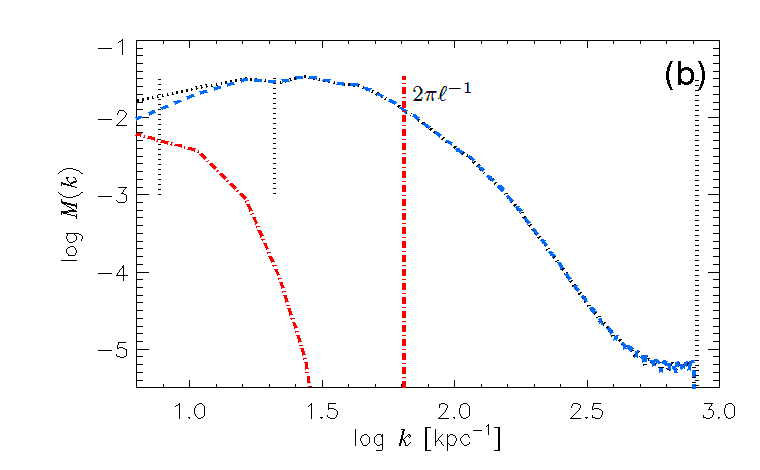}\vspace{-0.2cm}
  \caption{
  As Fig.~\ref{fig:ffta}, but for \textbf{(a)} $\ell=20\p$ and \textbf{(b)} 
  $\ell=100\p$.
  \label{fig:fftb}}
  \end{figure}

  Figure~\ref{fig:ffta} shows the magnetic power spectra of $\vect{B}$, 
  $\vect{B}_\ell$ and $\vect{b}{\frcorr{_\ell}}$.
  Note that $\ell$ is not located between the maxima in the power
  spectra of $\vect{B}_\ell$ and $\vect{b}{\frcorr{_\ell}}$; 
  in fact, the spectral density of the mean field is negligible for 
  $k\simeq2\pi\ell^{-1}$. 
  This can be understood from the transform of the kernel $G_\ell(\vect{x})$,
  i.e.\ $\widehat{G}_{\ell}(\vect{k})=\exp(-\ell^2 \vect{k}^2/2)$;
  this kernel would divide a purely sinusoidal field equally into the mean and
  random parts at the wavelength $\lambda_{\rm eq}=\sqrt{2/\ln 2}\,~\pi\ell$.
  For $\ell=50\pc$, $\lambda_{\rm eq}=0.27\kpc$,
  and the latter figure is a better guide to the expected separation scale.
  The separation of scales is immediately apparent in the spectra of the mean 
  and random fields, $M_\ell(k)$ and $M_b(k)$, with the former having a broad 
  absolute maximum at
  about $0.56\kpc$ and the latter, a broad
  maximum near $0.2\kpc$. 
The effective separation scale $\lambda{\frcorr{_{\rm eq}}}\approx0.48\kpc$ 
  can be identified where $M_\ell(k)=M_b(k)$, i.e. where
  the curves cross at $\log k\simeq1.1$.     
  {\frcorr{$\lambda>\lambda_{\rm eq}$ mainly characterize the mean and 
  $\lambda<\lambda_{\rm eq}$ the random field.}}

  The integral scales of $\vect{B}_\ell$ and $\vect{b}{\frcorr{_\ell}}$ can be obtained from
  their spectral densities as
  \begin{equation}
  \corr_\ell=\frac{\pi}{2}\,
  {\int_{{2\pi}/{D}}^{{\pi}/{\Delta}} k^{-1}M(k)\,\dd k} \left[
  {\int_{{2\pi}/{D}}^{{\pi}/{\Delta}} M(k)\,\dd k}\right]^{-1},
  \label{Bcorr}
  \end{equation}
  where $\Delta$ is the numerical grid separation and $D$ is the size of the 
  domain \citep[Section 12.1][]{MY07}. 
  This yields $\corr_\ell\simeq0.67\kpc$ and $\corr_b\simeq0.28\kpc$ for
  $\vect{B}_\ell$ and $\vect{b}{\frcorr{_\ell}}$, respectively (Fig.~\ref{fig:ffta}).

  As the magnetic field strength grows, the magnitudes of the spectral
  densities change but the characteristic wavenumbers vary rather weakly. 
  $\corr_\ell$ remains close to $0.7\kpc$ throughout the
  kinematic phase, increasing only beyond 1.1\,Gyr to about $0.9\kpc$. 
  $\corr_b$ increases from $0.23$ to $0.28\kpc$ during the kinematic phase,
  but rises to $0.4\kpc$ after the system saturates.
  The stability of $\corr_\ell$ is consistent with 
the eigenmode  
  amplification of the 
  mean magnetic field as expected for a kinematic dynamo, and supports 
  $\ell\approx50\p$ as a reasonable choice of the averaging scale.

  Figure~\ref{fig:fftb} presents magnetic energy spectra 
obtained  
  using
  $\ell=20\p$   and $100\p$.
  In the 
former, the effective separation scale $\lambda{\frcorr{_{\rm eq}}}\approx0.16\kpc$ is less
  than $\corr_b\simeq0.21\kpc$.  
  This is inconsistent, implying that energy at scales about $\corr_b$ lies
  predominantly within $\vect{B}_\ell$. 
  For $\ell=100\p$, 
$\lambda{\frcorr{_{\rm eq}}}\geq1\kpc$ is greater than $\corr_\ell\simeq0.81\kpc$,
  which is also inconsistent.
  Significantly, applying $\ell=50\p$ satisfies $\corr_b<\lambda<\corr_\ell$.
  Hence the scale $\lambda$, at which the dominant energy contribution
  switches between the mean and fluctuating parts, 
  is here consistent only with $\ell\simeq50\pc$.

  \begin{figure}
  \centering\vspace{-0.2cm}
  \includegraphics[width=0.85\linewidth]{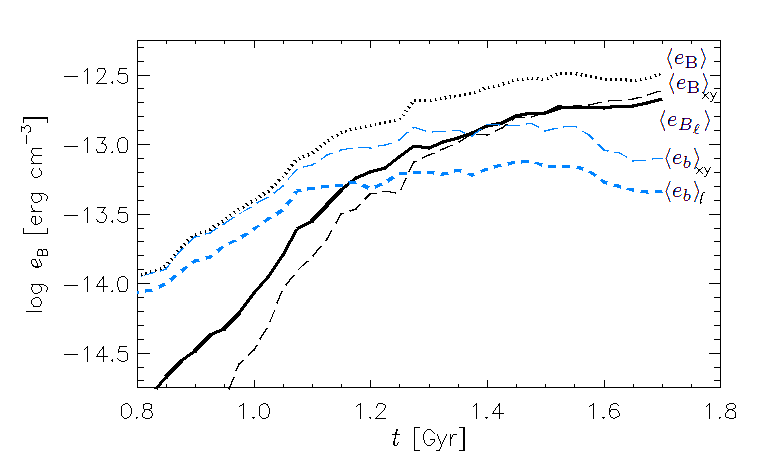}\vspace{-0.2cm}
  \caption{
  Evolution of magnetic energy densities, averaged for full domain:
  total magnetic field (black, dotted), mean (black, solid), and random 
  (blue, dashed) obtained from 
Gaussian smoothing
   with $\ell=50\p$.
  Energy densities of mean and random magnetic fields
  obtained from horizontal averaging are shown long-dashed (black and blue,
  respectively).
  \label{fig:bvt}}
  \end{figure}

  As well as different spatial scales, the mean and random magnetic field
  energies have different exponential growth rates (Fig.~\ref{fig:bvt}), given
  in Table~\ref{tab:fit}.
  Results from horizontal averaging are also given; they differ 
from those obtained with Gaussian smoothing,
  especially for the mean field.
  The growth rate $\Gamma$ of the mean-field energy is controlled by the shear
  rate, mean helicity of the random flow, and the turbulent magnetic 
  diffusivity. 
  Any mean magnetic field is accompanied by a random field, which is part of
  the mean-field dynamo mechanism (as any large-scale magnetic
  field is tangled by the random flow); 
  the energy of this part of the small-scale field should grow at the same rate
  $\Gamma$ as the mean energy.
  The fluctuation dynamo produces another part of the random field whose energy
  growth rate depends on the turbulent kinematic time scale $l/u$, and the
  magnetic Reynolds and Mach numbers.
  The difference between $\Gamma$ and $\gamma$ 
obtained suggests
  that both 
  mean-field and fluctuation dynamos are present
in our model.
  The growth rate of the mean magnetic energy is roughly double that of 
  the random field for both types of averaging.
  This is opposite to what is usually expected, plausibly because of the
inhibition of the fluctuation dynamo by the  
  strongly compressible nature of the flow,
  {\frcorr{and by the low Reynolds numbers available at this resolution.
  We would expect the fluctuation dynamo to be stronger with more realistic
  Reynolds numbers}}.

  \begin{table}
  \caption{\label{tab:fit}
  Exponential growth rates of energy in the mean and random 
  magnetic fields, $\Gamma$ and $\gamma$, respectively, 
  with associated values of reduced $\chi^2$. 
  From exponential fits to the corresponding 
  curves in Fig.~\ref{fig:bvt}, for $0.8<t<1.05\Gyr$.
  (Growth for $t<0.8\Gyr$ is similar to this interval
  -- see Fig.~\ref{fig:bbelog}.)
  }
  \centering
    \begin{tabular}{lcccc}
  \hline
                        &$\Gamma$      &$\chi^2$   &$\gamma$         &$\chi^2$ \\
                        &[Gyr$^{-1}$]  &           &[Gyr$^{-1}$]     &         \\
  \hline 
Gaussian smoothing   	&$10.9$        &$1.00$     &5.5              &$1.15$   \\
  Horizontal averaging	&$13.6$        &$0.25$     &$6.2$            &$0.25$   \\
  \hline
    \end{tabular}
  \end{table}

  \section{Discussion} \label{sect:conc}

  The approach used here to identify the appropriate averaging length $\ell$
  (and thus the effective separation scale $\lambda$) is simple (but 
  not oversimplified); $\ell$ can in fact depend on position, and it can remain 
  constant in time only at the kinematic stage of all the dynamo effects 
  involved.
  Wavelet filtering may prove to be more efficient than Gaussian smoothing in
  assessing the 
  variations of $\ell$.

  At the kinematic stage, the spectral maximum of the mean field is
  already close to the size of the computational domain, and it cannot be
  excluded that the latter is too small to accommodate the most
  rapidly growing dynamo mode.  These results should therefore be
  considered as preliminary with respect to the mean field;
  simulations in a bigger domain are clearly needed.
  
  The physically motivated averaging procedure used here, producing a mean
  field with three-dimensional structure, may facilitate fruitful 
  comparison of numerical simulations 
  with theory and observations.
  Although 
Gaussian smoothing
  does not obey all the Reynolds rules, 
  it is possible that a consistent mean-field theory can be developed,
  e.g.\ in the framework of the
  $\tau$-approximation \citep[see e.g.][and references therein]{BS05}. 
  This approach does not rely upon solving the equations for the
  fluctuating fields, and hence only requires the linearity
  of the average and its commutation with the derivatives.
  The properly isolated mean field is likely to exhibit different spatial and
  temporal behaviour than the lower-dimensional magnetic field obtained by 
  two-dimensional averaging. 

  \section*{Acknowledgments}
  We thank G.~L.~Eyink, A.~Brandenburg, M.~Rheinhardt,
  K.~Subramanian and E.~Blackman for discussions of
  averaging procedures.  Support: Grand Challenge project SNDYN,
  CSC-IT Center for Science Ltd., Finland; HPC-EUROPA2 Proj No.~228398;
  Academy of Finland Proj 218159 and 141017;
  Leverhulme Trust Research Grant RPG-097;
  STFC Grant F003080; UKMHD Consortium.

  \bibliographystyle{mn2e}      
  \bibliography{fred_mnras}
  \label{lastpage}
\end{document}